\documentclass[5p,
sort&compress
]{elsarticle}

\usepackage{hyperref}
\usepackage[utf8]{inputenc}
\usepackage{graphicx} 		
\usepackage{amsmath}  		
\usepackage{bm}
\usepackage{placeins}
\usepackage{braket}
\usepackage{gensymb}
\usepackage{subcaption}
\usepackage{booktabs}
\usepackage[capitalise]{cleveref}

\newcommand{\carbon}{$^{13}$C\:}
\newcommand{\proton}{$^{1}$H\:}
\newcommand{\JoneCH}{$^{1}\!J_{\mathrm{CH}}$\:}
\newcommand{\JtwoCH}{$^{2}\!J_{\mathrm{CH}}$\:}
\newcommand{\JoneCC}{$^{1}\!J_{\mathrm{CC}}$\:}
\newcommand{\JthreeHH}{$^{3}\!J_{\mathrm{HH}}$\:}
\newcommand{\Rb}{$^{87}$Rb\:}

\begin{document}
\emergencystretch 2em

\title{Two-Dimensional Single- and Multiple-Quantum Correlation Spectroscopy in Zero-Field Nuclear Magnetic Resonance}

\author[1]{Tobias~F.~Sjolander}
\author[2]{John~W.~Blanchard\corref{cor1}}
	\cortext[cor1]{Corresponding author: blanchard@uni-mainz.de}
\author[2,3,4]{Dmitry~Budker}
\author[1,5]{Alexander~Pines}
\address[1]{Department of Chemistry, University of California at Berkeley, CA, 94720 USA}
\address[2]{Helmholtz-Institut Mainz, GSI Helmholtzzentrum f{\"u}r Schwerionenforschung, 55128 Mainz, Germany}
\address[3]{Johannes Gutenberg-Universit{\"a}t  Mainz, 55099 Mainz, Germany}
\address[4]{Department of Physics, University of California, Berkeley, CA 94720-7300 USA}
\address[5]{Materials Science Division, Lawrence Berkeley National Laboratory, Berkeley, CA, 94720 USA}

\begin{abstract}
We present 
single- and multiple-quantum correlation 
$J$-spectroscopy detected in zero ($<\!\!1$~$\mu$G) magnetic field using a \Rb vapor-cell magnetometer. 
At zero field the spectrum of ethanol appears as a mixture of \carbon isotopomers, and correlation spectroscopy is useful in separating the two composite spectra. 
We also identify and observe the zero-field equivalent of a double-quantum transition in ${}^{13}$C$_2$-acetic acid, and show that such transitions are of use in spectral assignment. 
Two-dimensional spectroscopy further improves the high resolution attained in zero-field NMR since selection rules on the coherence-transfer pathways allow for the separation of otherwise overlapping resonances into distinct cross-peaks.
\end{abstract}


\maketitle

\section*{Introduction}
Nuclear magnetic resonance (NMR) experiments performed and directly detected in fields $<1 \,\mu$G \cite{Bernarding2006,Savukov2005,Ledbetter2008,Tayler2017} promise certain advantages over more conventional forms of NMR: the ability to perform ultra-high resolution spectroscopy without expensive superconducting magnets, and without the complications for spectral assignment often arising in intermediate fields due to strong-coupling effects \cite{Appelt2010}. Further, the absence of magnetic fields equalizes the Larmor frequencies (by setting them to zero) of distinct spin species, thereby allowing the study of physics not accessible at high fields. Examples include nuclear spin-singlet states formed by different nuclides \cite{Emondts2014a} and observation of terms in the nuclear spin-coupling Hamiltonian that are truncated at high field by large chemical-shift differences between nuclei. Such `non-secular' terms may be interesting for characterization of ordered materials \cite{Blanchard2015a}, and have been proposed as a means to detect both chirality \cite{King2016b} and molecular parity non-conservation using NMR \cite{JPNC}. Zero-Field NMR has also been explored as a test-bed for quantum-simulation \cite{Jiang2017}, offering an attractive combination of long ($>$10 s \cite{Emondts2014a}) coherence times and short ($<$ 1 $\mu$s \cite{Llor1995b}) control times.

Zero-field NMR experiments monitor the evolution of groups of coupled spins in the absence of an externally applied static magnetic field \cite{Thayer1987a,Blanchard2016}. The characteristic frequencies of the coherent evolution are determined by the spin-spin coupling Hamiltonian, which under rapid motional averaging is given by 
\begin{equation}
H_J = 2\pi\sum_{i>j}J_{ij}\bm{I}_i\cdot\bm{I}_j,
\label{eq:Hj}
\end{equation}
where the $J_{ij}$ are coupling constants (we write the $2\pi$ factor because $J_{ij}$ are in customarily given in Hz), the $\bm{I}_i$ are angular-momentum operators for each spin, and we have set $\hbar=1$. 
The result of such an experiment is known as a zero-field $J$-spectrum, which provides accurate fingerprints of different chemical species, despite the absence of any chemical-shift information \cite{Blanchard2013,Theis2013}. 
It is a feature of $J$-spectroscopy that only heteronuclear spin systems yield directly observable spectra \cite{Sjolander2017a}. 
This means that several common \proton containing solvents give no signal background, obviating the need for deuterated solvents. 
However, it also means that, since \carbon is only 1\% naturally abundant, the observed spectra of \proton-\carbon systems featuring more than one carbon atom at natural abundance are superpositions of contributions from the different possible \carbon isotopomers. Further complicating matters is the fact that the one-bond $J$-couplings of many organic molecules are of order $\sim$100\,Hz, which means that the peaks are spread out over only a few 100s of Hz in frequency space. 
While some molecules give $J$-spectra with peaks as narrow as 20 mHz \cite{Blanchard2013}, many other molecules do not. 
Additionally, spectral complexity increases rapidly with spin-system size. Taken together, these factors often lead to partially resolved or overlapping peaks, which complicates assignment. 

Meanwhile, the development of two-dimensional spectroscopy \cite{Jeener1979,Morris1986} is a major reason behind the analytical power of NMR. At a minimum, 2D experiments increase signal dispersion, thereby allowing the resolution of more crowded spectra. Additionally, many pulse sequences exist that enable the mapping of coupling networks, the simplification of spectral assignment, and structure elucidation \cite{Ernst1987}.

Multiple-quantum (MQ) NMR spectroscopy \cite{Hatanaka1975,Vega1976b,Norwood1992}, which in high field concerns transitions for which $|\Delta m| > 1$, where $m$ is the quantum number for the projection of the spin angular momentum on the field axis, has also found extensive use in NMR spectroscopy. In liquid-state analytical chemistry, multiple-quantum coherence filters combined with two-dimensional detection techniques provide one of the standard ways to map coupling networks \cite{Ernst1987}. MQ-spectroscopy also provides a means of simplifying the spectra of partially ordered systems, the smaller number of MQ peaks enables otherwise intractable spectra to be readily interpreted \cite{Warren1979,Warren1980}. In the solid state, MQ coherences may be used to monitor the growth of correlated spin clusters with applications to the investigation of the structure of amorphous solids \cite{Baum1985,Gleason1987}, and more recently in studies of many-body physics \cite{Alvarez2015,Wei2016}.

In this work we introduce two-dimensional correlation and MQ experiments in the context of liquid-state zero-field $J$-spectroscopy. 
Correlation spectroscopy is an attractive way to approach the problem of natural-abundance $J$-spectra containing contributions from different isotopomers, since coherence transfer between distinct molecules in liquids is not possible under normal circumstances. 
We show that the spectra from different \carbon isotopomers in ethanol may be separated from each other by observing the cross-peak pattern. 
The cross-peak pattern also simplifies spectral assignment and enables the distinction between otherwise overlapping resonances. 
We also show that [\carbon\!$_2$]-acetic acid supports the zero-field equivalent of a multiple-quantum transition, demonstrating for the first time the concept of Multiple-Quantum Zero- to Ultralow-Field (MQ-ZULF) NMR. 
Just as in high-field NMR, the number of MQ-ZULF transitions is significantly reduced compared to the number of single-quantum transitions, potentially leading to simpler, easier to interpret spectra. 
The selection rules governing the correlation pattern between single- and multiple-quantum coherences can be used to further simplify assignment. 
Finally, we note that the ability to perform 2D experiments with only one indirect dimension is a significant advantage offered by directly detected zero-field experiments, as opposed to using indirect detection in high field \cite{Suter1987a}, which requires two indirect dimensions.

\section*{Zero-Field Spin-State Manifolds}
Commonly, the most interesting features of two-dimensional spectroscopy are cross peaks due to coherence transfer from one transition to another. Therefore, selection rules that constrain the allowed pathways are important for the interpretation of 2D spectra. Here we show the origin of one such important constraint in zero-field $J$-spectroscopy.

In an isotropic liquid-state system at zero magnetic field, the nuclear spin eigenstates are also eigenstates of the total spin angular momentum operator $\bm{F}^2$ and its projection $F_\alpha$ on an arbitrary axis, and may conveniently be labeled with the quantum numbers $F$ and $m_F$ \cite{Butler2013}. This is most easily justified by noting that the nuclear spin Hamiltonian given in \cref{eq:Hj} is invariant with respect to rotations of the spin system and therefore must commute with $\bm{F}^2$ and $F_\alpha$ \cite{Butler2013}. However, for systems comprising more than two spin-1/2 nuclei, additional quantum numbers are necessary to fully define the zero-field eigenstates. It is particularly useful to consider the angular momentum of sets of magnetically equivalent spins \cite{Levitt2001}. `Equivalent' here denotes a set of indistinguishable spins that also share the same couplings to all other spins in the spin system. For a set of magnetically equivalent spins with total angular momentum $K$, there is no combination of pulses or evolution intervals that can break the equivalence and $\bm{K}^2$ commutes with all realizable effective Hamiltonians. Therefore, the presence of equivalent spins leads to selection rules in the zero-field spectra - the quantum numbers, $K$, associated with the total angular momentum of sets of equivalent spins must be conserved throughout any pulse-sequence.

\begin{figure}[t!]
	\includegraphics[width=\columnwidth]{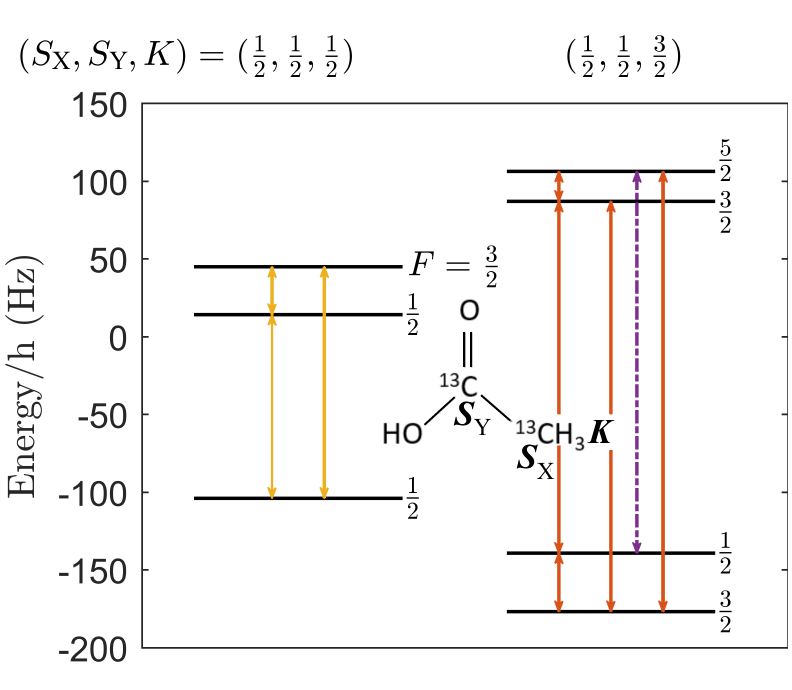}
	\caption{Energy levels of acetic acid may be grouped in two manifolds of different proton angular momentum ($K=1/2$, left, and $K=3/2$, right). Transitions and coherence transfer may only occur within each manifold. Magnetic dipole-allowed transitions have $\Delta F=\pm 1,0$ and are shown by solid lines. The dash/dot line marks a $\Delta F=\pm 2$ transition and the color coding and lines styles match \cref{fig:1DJspectra}}
	\label{fig:fullAcOH_Ediagram}
\end{figure}

\begin{figure}[t!]
	\includegraphics[width=\columnwidth]{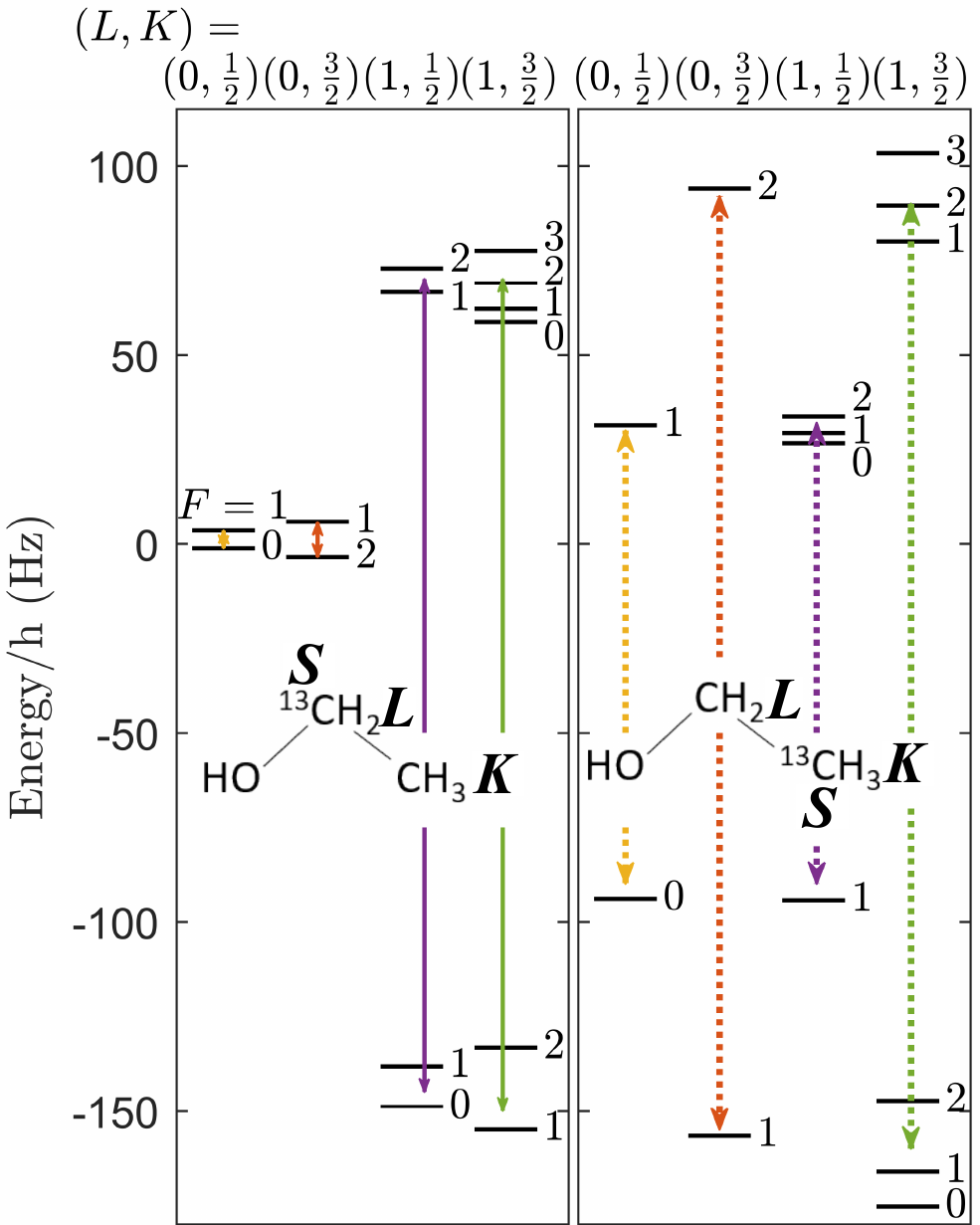}
	\caption{Energy levels of the two (singly labeled) \carbon isotopomers of ethanol support four distinct manifolds corresponding to different combinations of total proton angular momentum. For clarity only one transition per manifold is plotted and the quantum number $S=\frac{1}{2}$ for the \carbon spin has been omitted. Solid lines correspond to the 1-\carbon isotopomer and the dotted lines to the 2-\carbon isotopomer. The color coding and line styles match \cref{fig:1DJspectra}}
	\label{fig:ethanol_Ediagram}
\end{figure}

This selection rule reflects the existence of separate spin manifolds in zero field. 
We define such manifolds to be sets of energy levels which share the same combination of equivalent-spin quantum numbers. 
It follows from the above discussion that there can be no transitions of any kind between states belonging to different manifolds. 
As an example consider the zero-field energy level structure of \carbon\!$_2$-acetic acid (see Fig.\,\ref{fig:fullAcOH_Ediagram}), which, using the notation in \cite{Blanchard2016}, can be considered an (XA$_3$)Y spin system with the methyl (CH$_3$) protons being equivalent. 
The Hamiltonian for this spin system is $H = 2\pi(\:^{1}\!J_{\mathrm{CH}} \bm{S}_\mathrm{X}\cdot\bm{K} +\ ^{2}\!J_{\mathrm{CH}} \bm{S}_\mathrm{Y}\cdot\bm{K} +\ ^{1}\!J_{\mathrm{CC}} \bm{S}_\mathrm{X}\cdot\bm{S}_\mathrm{Y})$, where $\bm{K}$ and $\bm{S}_{\mathrm{X}/\mathrm{Y}}$ are angular momentum operators for the proton group and carbons respectively, and $\bm{K} = \sum_{i=1}^{3}\bm{I}_i$, where $\bm{I}_i$ refer to the proton spins. The three protons make up the only set of equivalent spins in the molecule and $K$ may take two values, 1/2 and 3/2. 
The allowed transitions may thus be assigned to two separate manifolds, as shown in the energy-level diagram in \cref{fig:fullAcOH_Ediagram}. 
There are actually two manifolds for which $K=1/2$, but they are degenerate, and we ignore this point for simplicity. 
There are no transitions between states of different $K$ and the two $K$ manifolds form entirely isolated spin systems. 
The principle extends readily to systems with more than one set of equivalent spins. 
For example, the Hamiltonian for the spin system of 1-\carbon-ethanol in zero-field is $H_{J} =2\pi(\ ^{1}\!J_{\mathrm{CH}}\bm{L}\cdot\bm{S} +\ ^{2}\!J_{\mathrm{CH}}\bm{K}\cdot\bm{S} +\ ^{3}\!J_{\mathrm{HH}}\bm{K}\cdot\bm{L})$, where $\bm{S}$ and $\bm{K}$ are defined as above and $\bm{L}$ is the operator for the angular momentum of the protons in the methylene (CH$_2$) group. 
1-\carbon-ethanol is an (XA$_2$)B$_3$ spin system, so $\bm{L}=\sum_{i} \bm{I}_{\bm{A},i}$ and $\bm{K}=\sum_{i} \bm{I}_{\bm{B},i}$.
The energy level diagram for this molecule, as well as the 2-\carbon-isotopomer is shown in \cref{fig:ethanol_Ediagram}. 
In both cases there are four spin-manifolds, corresponding to the four distinct combinations of $K$ and $L$. 

Since a spin-state labeled with a certain combination of conserved quantum numbers may not evolve or transform into a spin-state labeled with a different combination, it follows that in 2D zero-field spectroscopy we will not see cross-peaks between transitions belonging to different spin-manifolds. 

This phenomenon is  not unique to zero-field NMR; spin states belonging to different irreducible representations of a given permutation group do not mix or evolve into each other no matter the magnitude of the external field. For example, in high-field NMR a ${}^{13}$CH$_2$ group (where the protons are magnetically equivalent) gives only two peaks at the proton frequency, since transitions between the proton singlet and triplet states are forbidden.

\section*{Zero-Field Total Correlation Spectroscopy}

The building blocks of a 2D NMR experiment are the same for zero- and high-field experiments. A general schematic is shown in \cref{fig:pulseSequences}a. 
First, the desired coherences are prepared in an excitation step, described by the overall propagator $U_{exc}$; this is followed by free evolution during which these coherences acquire phase. 
Finally, a reconversion step, described by the propagator $U_{re}$, precedes the readout. 
In this work, the excitation and reconversion sequences are generated by one or several DC magnetic-field pulses around different axes.

The pulses are much stronger than any $J$-couplings for these systems, $|\gamma B| \gg |J|$, so to a good approximation each spin is independently rotated by an angle $\theta_i = \gamma_i Bt_p$ around the pulse axis, where $\gamma_i$ is the gyromagnetic ratio of the $i^{\mathrm{th}}$ spin, $B$ is the amplitude of the pulse, and $t_p$ is its duration. 
In this work, we use \proton + \carbon spin systems and all pulses are calibrated to effect a $\pi$ rotation of the \carbon spins, which means that the \proton spins are rotated by $\pi\times(\gamma_{^1\mathrm{H}}/\gamma_{^{13}\mathrm{C}})\approx 4\pi$. 
We take the detection axis to be the z-axis and assume that the initial state in all experiments is given by magnetization along the z-axis generated by adiabatic transport from a pre-polarizing magnetic field, giving the initial deviation density matrix $\rho(0) \propto \sum_i I_{z,i}$. 
This state commutes with $H$ and so does not evolve. 
Evolution may be initiated by changing the relative orientation of the proton and carbon spins, for example with a $\pi$ pulse on carbon in the x/y plane \cite{Blanchard2016}.

Reference \cite{Sjolander2016} introduced a zero-field `spin-tickling' method whereby assignment of resonances was simplified by monitoring the response of the spectrum to low-amplitude irradiation of selected transitions. 
We present a two-dimensional variant, where the complete spectral connectivity is established in one experiment. 
Here, $t_1$ evolution is initiated with a $\pi$ pulse on carbon along the x-axis followed by a second $\pi$ pulse and $t_2$ evolution (detection). 
The Fourier transform with respect to $t_1$ and $t_2$ gives a correlation spectrum between two spectral dimensions, F1 and F2, each of which corresponds to a zero-field $J$ spectrum. 
The sequence is summarized schematically in Fig.\ \ref{fig:pulseSequences}b. 
We refer to this type of experiment as ZF-TOCSY, so named because of the similarity to high-field Total Correlation Spectroscopy (TOCSY) \cite{Braunschweiler1983}, in which a spin-lock provides an effective strong-coupling condition during the isotropic mixing period.
The zero-field variant takes advantage of the fact that all nuclear spins are already strongly coupled at zero field, so no Hamiltonian engineering is required.

\section*{Zero-Field Multiple-Quantum Correlation Spectroscopy}
\begin{figure}[]
	\includegraphics[width=\columnwidth]{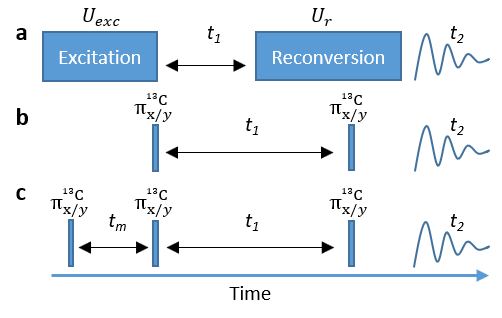}
	\caption{General scheme for 2D NMR spectroscopy and  the pulse sequences used in this work. (a) An excitation sequence prepares the desired coherences, this is followed by a period of free evolution and finally a reconversion sequence and detection. (b) The ZF-TOCSY sequence. (c) An example MQ-ZULF sequence.}
	\label{fig:pulseSequences}
\end{figure}
In high-field NMR spins are naturally quantized along the external magnetic field and states may be labeled with their projection, $m$, along that field. Directly observable single-quantum coherences have $\Delta m = \pm 1$, and indirectly observable multiple-quantum coherences are those for which $|\Delta m| > 1$. 
Conversely, in zero field the eigenstates are eigenstates of total angular momentum, labeled $F$. 
The observable in both high- and zero-field NMR experiments is total magnetization along some direction, conventionally taken to be the x-axis in high field and the z-axis in zero field. 
Since the sample magnetization is represented by a vector operator, it follows from the Wigner-Eckart theorem \cite{VMK} that the allowed transitions are those with $\Delta F = 0, \pm 1$. 
We suggest that the indirect observation of transitions for with $|\Delta F| > 1$ may serve as a zero-field analog to high-field MQ experiments. 

An example pulse sequence that can be used to observe such transitions is shown in \cref{fig:pulseSequences}c. 
In this sequence, which is modeled on the original MQ-excitation sequence, the unitary propagator for the excitation is given by $U_{exc} = P_x-U_{t_m}-P_x$, and the reconversion propagator is $U_{re} = P_x$. 
$P_x$ is the propagator for a $\pi$ pulse along $x$ on \carbon and $U_{t_m}$ the propagator for free evolution for a time $t_m$. 
Starting from single-spin order, the first pulse initiates evolution under the $J$-coupling Hamiltonian, which subsequently generates multiple-spin terms. 
The second pulse converts some of those terms into zero-field MQ-coherences. After $t_1$ evolution these terms are converted into observable magnetization with a readout pulse.

With the exception of the first pulse and the mixing period, this pulse sequence is identical to the ZF-TOCSY sequence and can therefore be expected to produce a similar spectrum. 
The difference is that the $t_1$ period contains both observable and $n>1$ coherences and we expect the resulting spectrum to look like a ZF-TOCSY spectrum, but with additional cross-peaks in F1 due to coherences evolving at frequencies corresponding to $n>1$ transitions during $t_1$. 

\begin{figure*}[t!]
	\centering
	\includegraphics[]{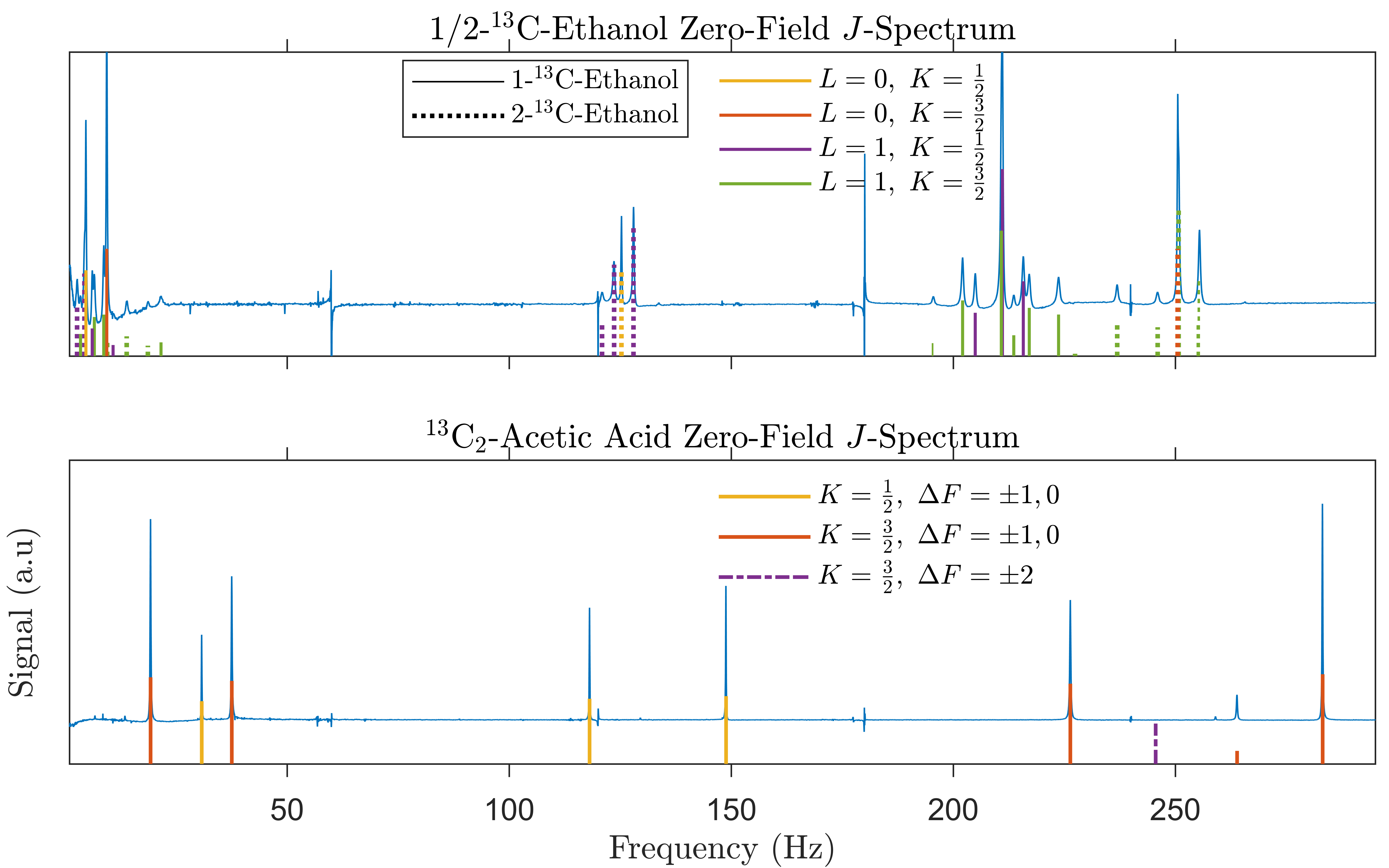}
	\caption{One dimensional zero-field $J$-spectra of ethanol and \carbon\!$_2$-acetic acid. The stick spectra correspond to the transition energies predicted by numerical diagonalization of the spin Hamiltonian. The transitions have been labeled and color-coded according to their energy-level manifold. For the ethanol spectrum, solid lines correspond to the 1-\carbon isotopomer and dotted lines to the 2-\carbon isotopomer. In the acetic acid spectrum the expected frequency of a $K$-conserving zero-field double-quantum transition is shown for reference.
	Note that narrow lines at multiples of 60\,Hz arise from power line noise.}
	\label{fig:1DJspectra} 
\end{figure*} 

The possible values of $n$ that can occur during $t_1$ depend on the spin topology. 
Given a spin system containing $k$ sets of equivalent spins we can write an expression for the highest order of zero-field quantum coherence supported by each energy-level manifold. 
Each manifold is labeled by $k$ quantum numbers and $n_{\mathrm{max}}$ for a particular manifold is given as twice the largest quantum number in the manifold, or twice the sum of the remaining ones, whichever number is smaller: 
\begin{equation}
n_{\mathrm{max}}=2\times\mathrm{Min}\lbrace \sum_{i=1}^{k-1}f_i\ ,\ f_{k} \rbrace,
\label{eq:nmax}
\end{equation}
where $f_i$ is the total angular momentum for each set of equivalent spins and the sum runs over the $k-1$ smallest quantum numbers in the manifold. 
$f_{k}$ is the largest quantum number in the manifold. 
As an example consider the energy level diagram of $^{13}$C$_2$-acetic acid given in Fig.\ \ref{fig:fullAcOH_Ediagram}. 
There are three groups of inequivalent spins, meaning $k=3$. The ($f_1 = 1/2, f_2 = 1/2, f_3=1/2$) manifold supports only $n=0,1$ transitions consistent with the fact that the largest quantum number, $f_3$, in the manifold is $1/2$, whereas the ($f_1 = 1/2, f_2 = 1/2, f_3 = 3/2$) manifold supports a $n=2$ transition, since $2\times (f_1+f_2) = 1$. 
The MQ-ZULF experiment thus provides direct information regarding the possible quantum numbers associated with a given resonance.

\Cref{eq:nmax} can be justified as follows: $n_{max}$ denotes the largest possible change in total angular momentum that may occur in a given spin-manifold, meaning we have $n_{max}=F_{max}-F_{min}$. 
Each $F$ value is the result of successively coupling all angular momenta $f_i$ in that manifold. 
$F_{max} = \sum_{i=1}^{k}f_i$, and $F_{min} = |f_k - \sum_{i=1}^{k-1}f_i|$. 
Depending on which of $f_k$ or $\sum_{i=1}^{k-1}f_i$ is bigger we obtain the two cases for \cref{eq:nmax}. 

\begin{figure*}[t!]
	\begin{subfigure}[b]{0.9\columnwidth}
		
		\includegraphics[]{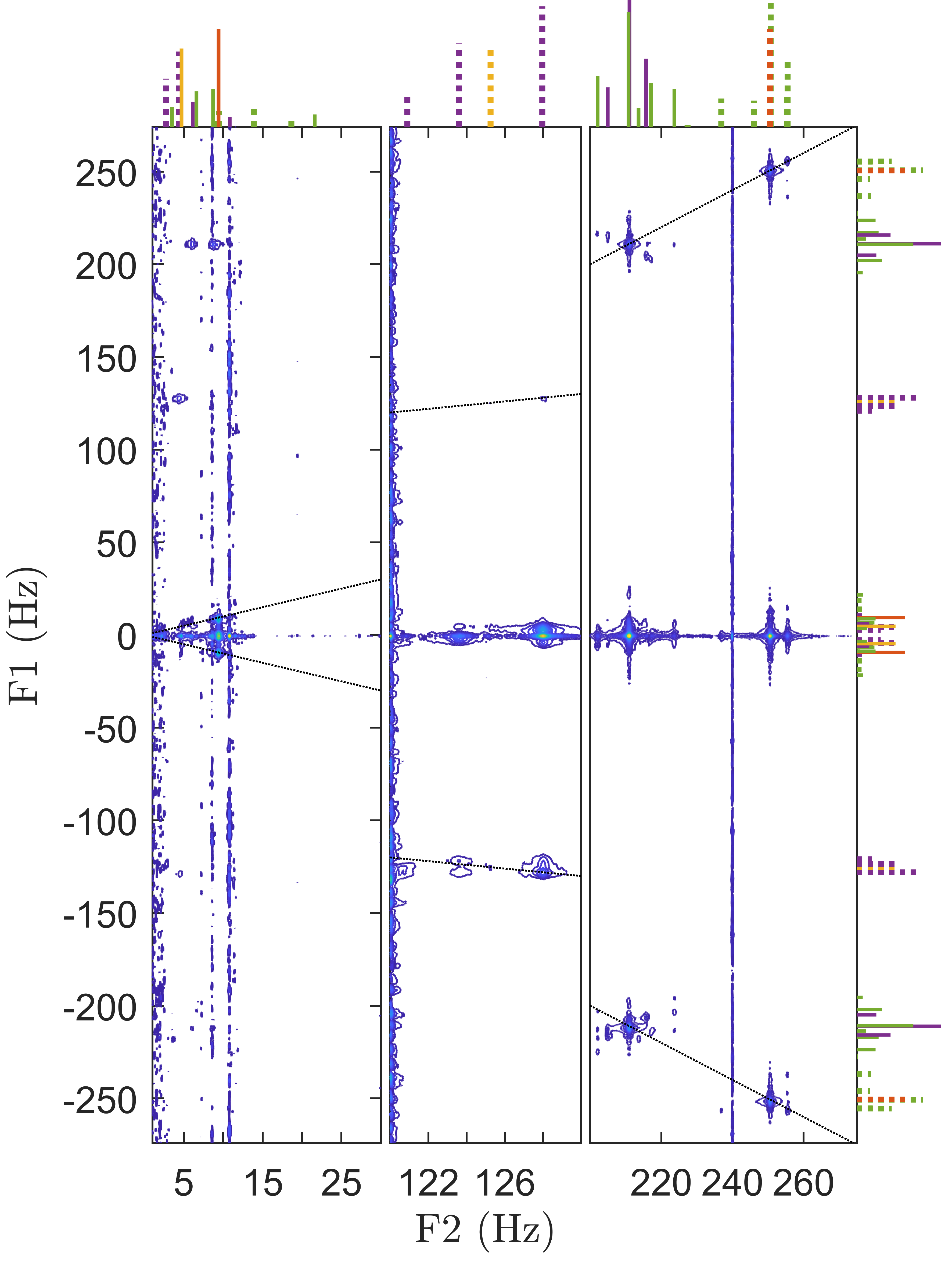}
		\caption{Ethanol Mixture}
		\label{fig:EthanolTOCSY}
	\end{subfigure}
	\hfill
	\begin{subfigure}[b]{0.9\columnwidth}
		
		\includegraphics[]{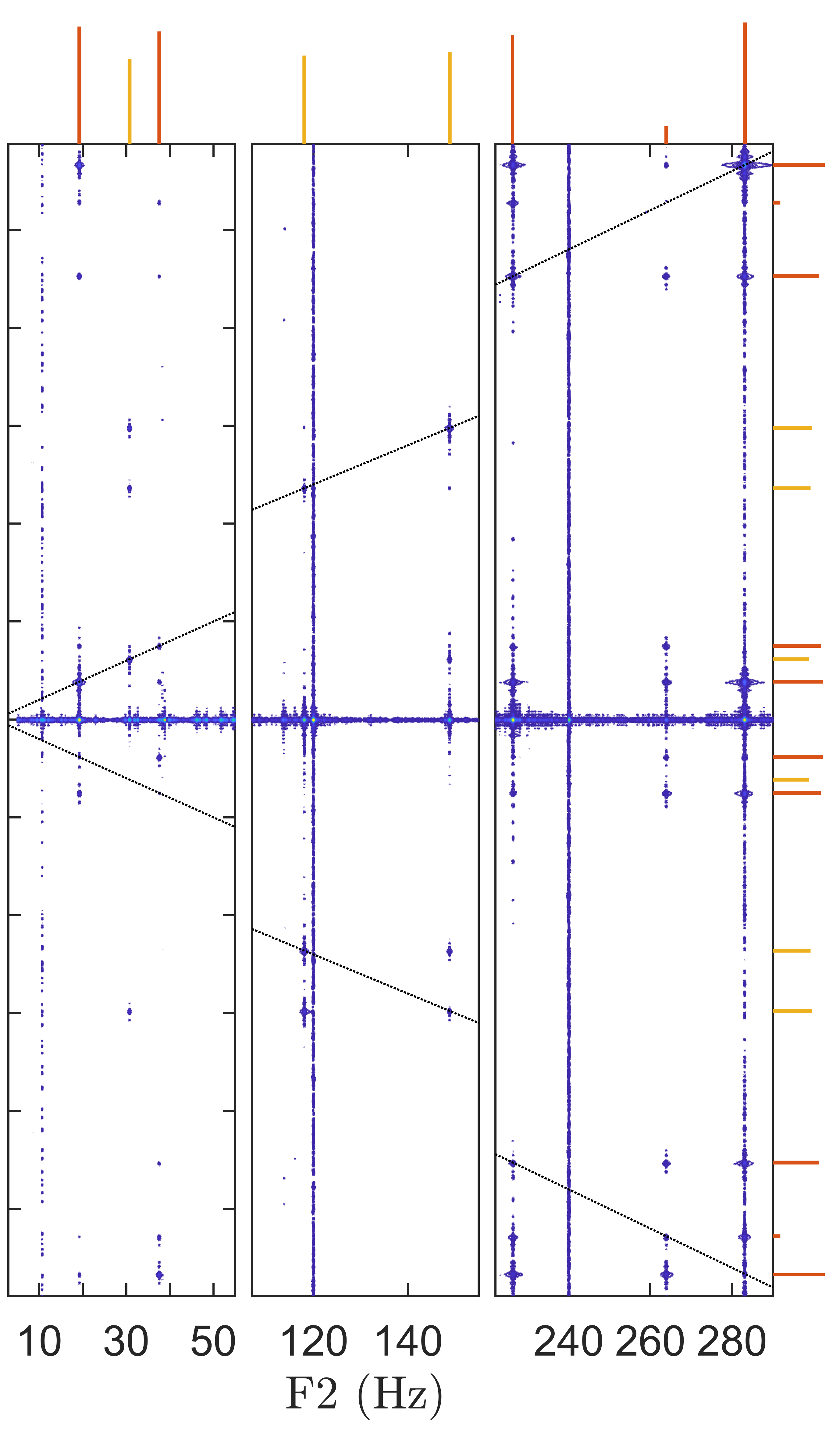}
		\caption{\carbon\!$_2$ Acetic Acid}
		\label{fig:fullAcOH_TOCSY}
	\end{subfigure}
	\caption{Zero-Field Total Correlation Spectra acquired using the protocol in Fig.\ \ref{fig:pulseSequences}b. The numerical peak assignments made in Fig.\ \ref{fig:1DJspectra} are plotted on both axes and the diagonal is shown as a dotted black line for reference.}
\end{figure*}

\section*{Methods}
We performed experimental demonstrations of the ZF-TOCSY and MQ-ZULF experiments using \carbon labeled isotopomers of acetic acid and ethanol. 
In the case of ethanol, we prepared a sample consisting of a mixture of the two singly labeled \carbon isotopomers.
The ratio of the isotopomers in this mixture was equal, as in natural samples. 
The acetic acid sample was doubly \carbon labeled in order to ensure the presence of a $K$-conserving double-quantum transition. 
The experiments were performed using a \Rb vapor-cell magnetometer operating in the Spin-Exchange 
Relaxation-Free (SERF) regime \cite{Kominis2003}, configured for use as an NMR spectrometer \cite{Tayler2017}. 
SERF magnetometers are DC magnetic field sensors, which allows us to directly monitor the low-frequency spin evolution in zero-field. 
All experiments were done using $\sim$80 $\mu$L samples in 5 mm outer diameter standard NMR tubes. 
The acetic acid sample contained only \carbon\!$_2$-acetic acid (residual singly labeled material is detectable, but this has negligible effect on the experiment) while the ethanol sample was made from $\sim$40 $\mu$L 1-\carbon-ethanol and $\sim$40 $\mu$L 2-\carbon-ethanol. 
The samples were prepolarized in a 2 T permanent magnet and shuttled pneumatically into a magnetically shielded region for zero-field evolution and detection. 
For the 2D experiments, the samples were re-polarized between every point in the indirect dimension.

\section*{Results and Discussion}
For reference, we first obtained 1D $J$ spectra of these molecules; the results are presented in \cref{fig:1DJspectra}. 
The spectra are the result of summing 2000 and 3000 transients, respectively, for the acetic acid and ethanol data. 
In both cases, each transient corresponds to 20 s of data acquisition. 
Numerical analysis is used to assign the eigenvalues of $\bm{K}^2$ and $\bm{L}^2$ for each transition.
Extracted $J$-coupling values are provided in the appendix, along with a discussion of further details of the spectra.
Simulated spectra based on the best-fit $J$ values, together with the $K$ and $L$ assignments are also shown in \cref{fig:1DJspectra}. 

\begin{figure}[t!]
	\centering

	\includegraphics[width=\columnwidth]{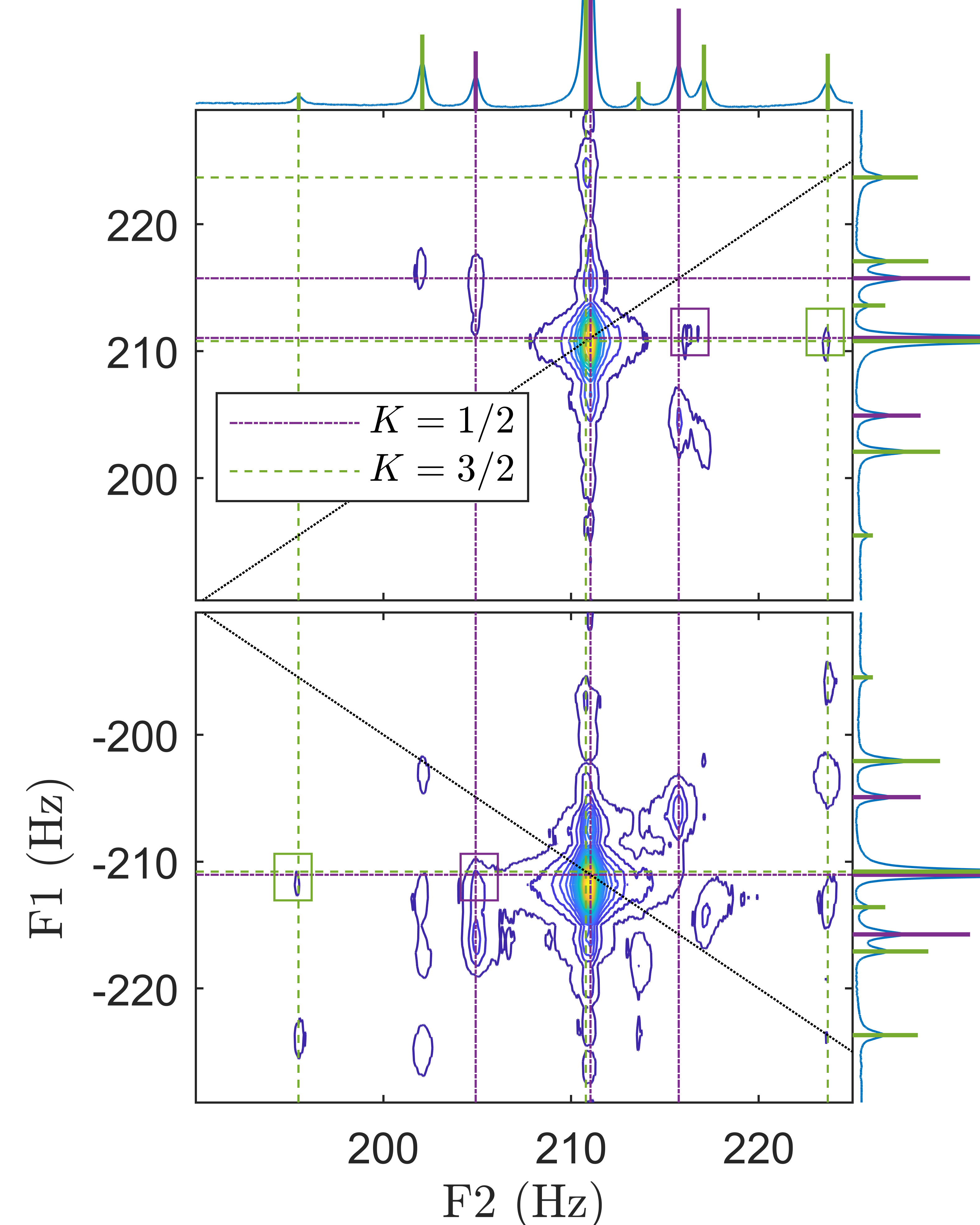}
	\caption{Detailed structure of the high-frequency multiplet in the ZF-TOCSY spectrum of 1-\carbon ethanol. All peaks in this multiplet have $L$=1. The dashed and dotted lines are drawn to guide the eye to particular cross-peaks (which are marked by boxes) which confirm that the high-intensity peak at 211\,Hz actually consists of two overlapping resonances which belong to different $K$-manifolds.}
	\label{fig:Zoomed}
\end{figure}

The 1D $J$-spectra may be interpreted a follows: 
In $^{13}$C$_2$-acetic acid, the dominant coupling is between a single spin-1/2 carbon coupled to a group of three equivalent protons with angular momentum $K$. 
An XA$_3$ spin system has two transition frequencies depending on the value of $K$, at $J$ and 2$J$, for $K$ = 1/2, and 3/2 respectively. 
Couplings to the second carbon result in the spectrum being made up of groups of transitions centered at those two frequencies plus additional peaks close to zero \cite{Theis2013}. 
In the experimental data we see two peaks in the 120-150\,Hz range, three peaks between 225\,Hz and 280\,Hz, and 3 peaks between 20 and 40\,Hz, while the best-fit value for $^{1}\!J_{\mathrm{CH}}$ was 129.504\,Hz. 
Careful inspection of the spectrum also reveals weak signals at 6.75, 13.5, 129.5 and 259\,Hz. 
These peaks can be assigned to residual 1-\carbon and 2-\carbon acetic acid in the sample. 
Both isotopomers would yield peaks at 1$J$ and 2$J$ on account of being XA$_3$ spin systems, and the corresponding coupling constants listed in the appendix in \cref{tab:exptJCouplingsEthanol} are consistent with the positions of the weak signals. 
Finally, the stick spectrum also shows the expected position of the $K$-conserving $\Delta F = \pm2$ transition, however since this transition does not correspond to oscillating magnetization it is not observed in the directly detected 1D data.

For both isotopomers of ethanol, the spectrum is to first order determined by a strong $^{1}\!J_{\mathrm{CH}}$ coupling \cite{Theis2013}, which sets up an initial splitting pattern that is further split by the weaker $^{2}\!J_{\mathrm{CH}}$ and $^{3}\!J_{\mathrm{HH}}$ couplings. 
An XA$_2$ spin system in zero-field gives a single peak at 3/2$J$ while an XA$_3$ system gives one peak at the coupling frequency and another at twice the coupling frequency. 
We can therefore identify the cluster of peaks at 210\,Hz with the 1-\carbon isotopomer and the peaks around 125\,Hz and 250\,Hz with the 2-\carbon isotopomer. 
This corresponds to a one-bond \carbon/\proton $J$-coupling constant of $\sim$140\,Hz when the \carbon label is on the methylene group, and $\sim$125\,Hz when the \carbon is on the methyl group. This is consistent with the results obtained by numerical fitting of the spectra. 
While this back-of-the-envelope interpretation gives approximate values for the 1-bond coupling constants, we note that without the aid of numerical simulations it would for example be difficult to say with any certainty where the spectrum of the 1-\carbon isotopomer ends and that of the 2-\carbon isotopomer begins. 
Without computer assistance it would also be challenging to distinguish the $K=1/2$ peaks from the $K=3/2$ peaks in the 3$J$/2 multiplet associated with the 1-\carbon isotopomer. 
However, not all spin systems are small enough that their Hamiltonians are readily diagonalizable, and in chemical analysis the exact coupling topology is not always known in advance. 
Here we dmonstrate how to overcome these problems using 2D spectroscopy.

\begin{figure*}[t!]
	\begin{subfigure}[b]{0.9\columnwidth}
		
		\includegraphics[]{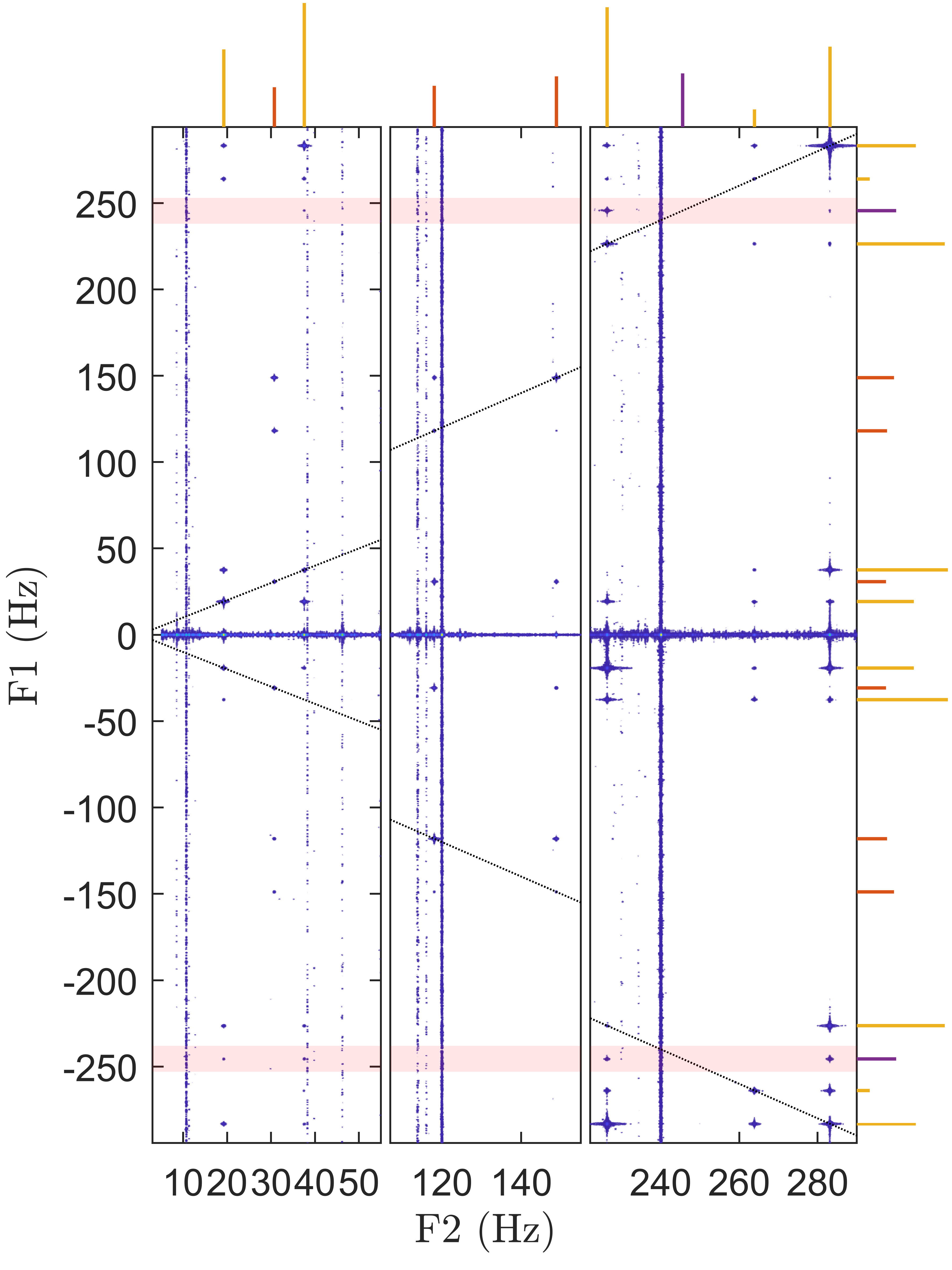}
		\caption{}
		\label{fig:fullAcOH_MultiPolar2}
	\end{subfigure}
	\hfill
	\begin{subfigure}[b]{0.9\columnwidth}
	
		\includegraphics[]{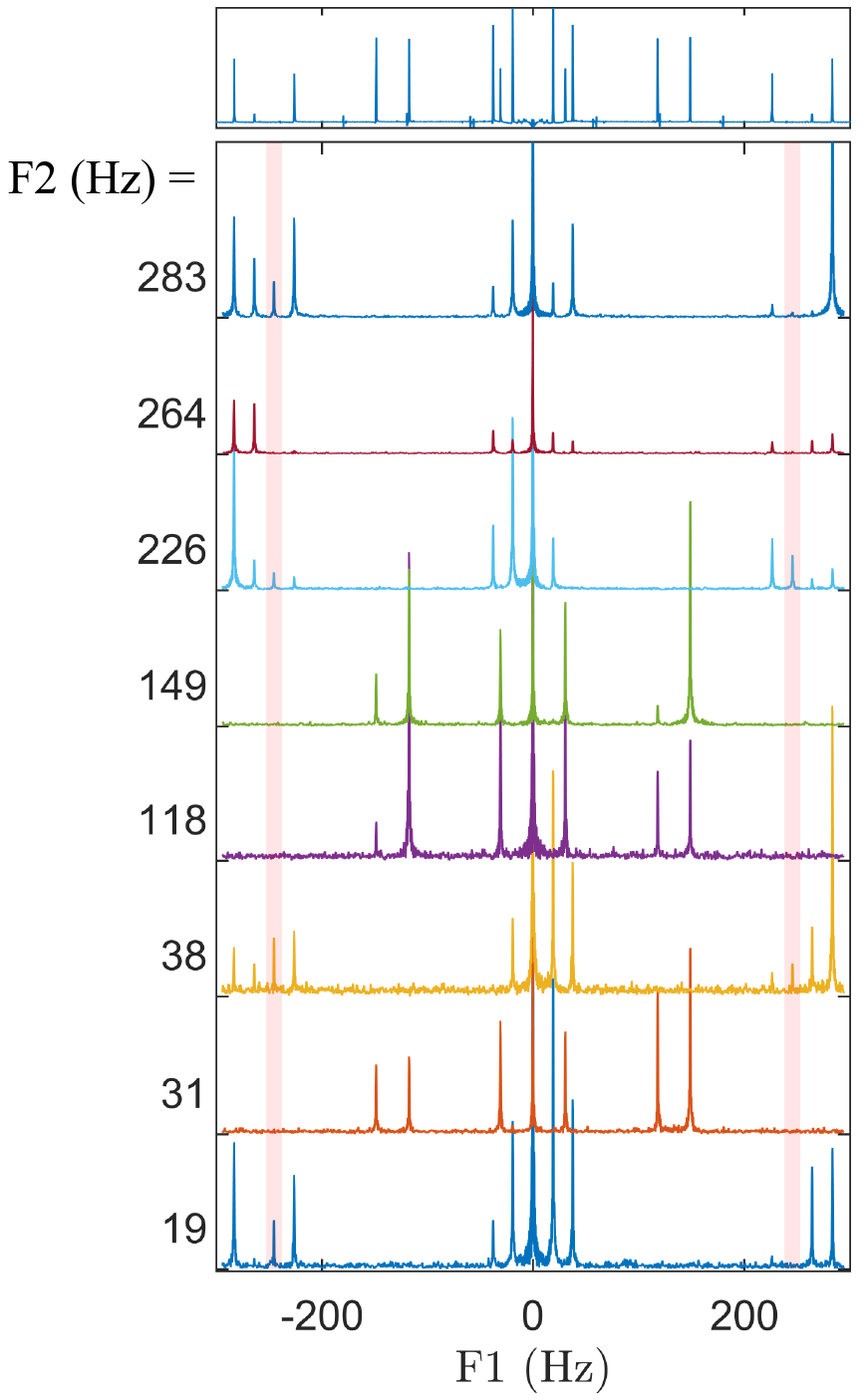}
		\caption{}
		\label{fig:fullAcOH_MultiPolar}
	\end{subfigure}
	\caption{MQ-ZF spectra of \carbon\!$_2$ acetic acid. (a) 2D Single/multiple-quantum correlation spectrum. F1 corresponds to the indirectly detected dimension, and the numerical peak assignments of \cref{fig:1DJspectra}b are plotted on both axes for reference. (b) Slices through the 2D spectrum in \cref{fig:fullAcOH_MultiPolar2}. The x-axis corresponds to F1, and each spectrum is the projection of the data around a $\pm$ 0.5\,Hz slice at the designated frequencies in the directly detected dimension (F2). The pink bands indicate $\pm$ 245\,Hz, the expected frequency of the double-quantum transition. Note that for convenience the spectra are evenly spaced in the vertical dimension. The top panel shows the directly detected 1D $J$-spectrum.}
\end{figure*}

Figures \ref{fig:EthanolTOCSY} and \ref{fig:fullAcOH_TOCSY} show the ZF-TOCSY spectra of the ethanol mixture and the acetic acid sample acquired using the protocol in \cref{fig:pulseSequences}b. 
Note that the entire spectral window (both positive and negative frequencies) is displayed in F1, whereas only the positive part of the F2 axis is shown. 
This is because the recorded signal is purely real, and its Fourier Transform is therefore conjugate symmetric. 
The spectra are displayed in magnitude mode, so to display all the spectral information it is therefore only necessary to plot two of the four quadrants of the Fourier transformed data. 
Overtones of the power-line noise lead to vertical streaks in the data at multiples of 60\,Hz and a climate-control fan causes building vibrations at $\sim$ 10.7\,Hz. 
We do not know the origin of the feature near 8.7\,Hz. 
The positions of the transitions based on the fitted $J$-coupling constants are plotted on the axes for reference. 

The spectra contain cross-peaks between transitions belonging to the same spin-state manifolds. 
Meanwhile, there is no coherence transfer, and therefore no cross-peaks, between either the two isotopomers or peaks corresponding to the same isotopomer but different combinations of $K$ and $L$. 
This allows us to confirm the numerical assignment of the peaks made in \cref{fig:1DJspectra}. 
For example the peak at $\sim$238\,Hz in the ethanol spectrum in \cref{fig:EthanolTOCSY} clearly correlates to the 2$J$ multiplet at around 250\,Hz and therefore belongs to the 2-\carbon isotopomer. 
Similarly the peak at $\sim$125\,Hz does not give cross peaks with any of the three other peaks in the 1$J$ multiplet, and it therefore must belong to a separate spin-manifold, consistent with the numerical assignment of the 125\,Hz peak to $L=0$ and the three surrounding peaks to $L=1$. 
With the same reasoning the ZF-TOCSY spectrum of acetic acid in \cref{fig:fullAcOH_TOCSY} can be used to distinguish between the three lower-frequency peaks of the acetic acid $J$-spectrum, and confirm the numerical assignment of the peak at 31\,Hz to the same spin-manifold as the peaks at 119 and 149\,Hz. 
These assignments can all be made without numerical diagonalization.

As a further example of the use of 2D spectroscopy, consider the high-frequency multiplet of the 1-\carbon isotopomer (at around 3/2$J$ in both F1 and F2) shown in more detail in \cref{fig:Zoomed}. 
Fitting of the 1D data reveals that the high-intensity peak at $\sim$211\,Hz actually consists of two overlapping resonances with different $K$ quantum numbers. 
This is confirmed by the cross-peak pattern, as \cref{fig:Zoomed} shows that the 211\,Hz peak correlates to both $K=1/2$ and $K=3/2$ resonances in F1. 
In this case 2D spectroscopy allows us to distinguish overlapping resonances by correlating them to distinct, well separated, peaks.

The ZF-TOCSY experiment relies on similar physics to high-field Total Correlation Spectroscopy (TOCSY) \cite{Braunschweiler1983}, but does not require a mixing time. 
This follows from the fact that the zero-field free-evolution Hamiltonian already provides strong coupling for all spin pairs (both homo- and heteronuclear) and thus allows complete coherence transfer throughout the molecule. 
The same fact also means that F$_1$ and F$_2$ do not correspond to individual-spin transitions but rather to zero-field $J$-spectra. 
Previous 2D experiments with direct detection were either performed with different effective Hamiltonians during the two evolution intervals \cite{Sjolander2017a}, or in the presence of a magnetic field such that the Larmor frequencies and $J$-coupling frequencies are approximately equal \cite{Shim2014a}. 
These cases lead to significantly different cross-peak patters.

As mentioned above, inspection of the energy-level diagram of ${}^{13}$C$_2$-acetic acid shown in \cref{fig:fullAcOH_Ediagram} reveals a $K$ conserving, and therefore potentially observable, $\Delta F = \pm 2$ transition at $\sim$245\,Hz. 
The results of a MQ-ZULF experiment designed to observe this transition, in spite of it not corresponding to oscillating magnetization, is shown in \cref{fig:fullAcOH_MultiPolar2}. 
The data were acquired using the pulse sequence in \cref{fig:pulseSequences}c. 
The cross-peak pattern is mostly the same as in the ZF-TOCSY spectrum in \cref{fig:fullAcOH_TOCSY}, however there is one additional peak in F1 at 245\,Hz (highlighted pink band) corresponding to oscillations during $t_1$ at the frequency of the $\Delta F = \pm2$ transition. 
According to \cref{eq:nmax} only the $K=3/2$ manifold supports such a transition, and it shows up only as cross-peaks to the transitions at 19\,Hz, 38\,Hz, 225\,Hz, 265\,Hz, and 283\,Hz, thus confirming the numerical assignment made in \cref{fig:1DJspectra}b of those peaks to the $K=3/2$ manifold. 
This is perhaps seen more clearly in \cref{fig:fullAcOH_MultiPolar}, which shows slices through the indirect (F1) dimension taken at the positions of the peaks in 1D spectrum. 
The MQ-resonance (the expected position is indicated with a pale band) clearly shows up in the indirectly detected data, but only in those spectra that are read out at the frequencies of the $K=3/2$ transitions.

We note that the initial density matrix also contains terms proportional to scalar order $\propto\sum_{ij}\bm{I}_i\cdot\bm{I}_j$ \cite{Emondts2014a,Theis2012}, which already contains two-spin terms at time zero. 
This could therefore be turned into a two-pulse experiment if pulses along $z$ are used instead, since such pulses access scalar spin-order instead of vector spin-order \cite{Emondts2014a}.

\section*{Conclusions}
We have shown that direct detection using magnetometers facilitates 2D zero-field NMR correlation spectroscopy, and how such techniques simplify assignment of crowded $J$-spectra. 
The complete coherence transfer enabled by the isotropic zero-field Hamiltonian ensures that cross peaks in the ZF-TOCSY spectrum are seen between all peaks belonging to the same spin-manifold. 
Consequently, ZF-TOCSY may be used not only to distinguish between different molecules, or different \carbon isotopomers of the same molecule, but also to facilitate zero-field spectral assignment of a given isotopomer by providing a way to determine if two transitions occur within the same angular-momentum manifold. 
Additionally, 2D-spectroscopy increases the maximum attainable spectral resolution by introducing a second spreading parameter in the spectrum. In particular, the ability to resolve otherwise overlapping peaks significantly increases the power of zero-field NMR for chemical fingerprinting, beyond what can already be obtained with the narrow linewidths associated with high-homogeneity zero-field environments.

We have also introduced the concept of multiple-quantum transitions in zero- to ultralow-field NMR (MQ-ZULF) and shown that such transitions may only belong to particular spin-manifolds. 
Therefore, by observing which peaks correlate to a multiple-quantum transition, one can assign quantum numbers to those peaks. 
Filtering zero-field coherences \cite{Ernst1987,Shaka1983} based on $\Delta F$ would allow for further simplification of spectra -- this remains an outstanding experimental challenge to be addressed in future work (preliminary efforts are discussed in the appendix).

\section*{Acknowledgements}
This work was supported by the National Science Foundation under award CHE-1709944.
The authors thank Rom{\'a}n Picazo Frutos for helpful comments on the manuscript.

\bibliography{library}

\section*{Appendix: Detailed Analysis of 1D $J$ Spectra}

The $J$-coupling constants for the two ${}^{13}$C isotopomers of ethanol and ${}^{13}$C$_2$-acetic acid were obtained from the spectra by a numerical fitting procedure, where the data are matched to the spectra predicted by numerical diagonalization of the spin Hamiltonians given in the main text. 
The results are given in \cref{tab:exptJCouplingsEthanol}. 

In the case of ethanol, we find that the measured 3-bond \proton-\proton coupling constant is different between the two \carbon-isotopomers. 
Since both isotopomers were present in the same NMR tube at the same time during the experiment most potential sources of systematic error may be discarded and we assign the measured difference in the \proton-\proton couplings to a secondary isotope shift to the $J$-coupling constant \cite{Chertkov1983,Sergeyev1990,Wilzewski2017}. 
Such effects have been of interest since they allow for a check on quantum chemical calculations and our understanding of bonding. 

We also note that the measured linewidths are significantly larger for ethanol than for acetic acid, and both spectra are in turn significantly broader than what was previously recorded for $J$-spectra of systems without labile protons \cite{Blanchard2013}. 
We note that in strongly coupled systems, chemical exchange will lead to decreased coherence times \cite{Barskiy2019}.
Polarization is exchanged throughout the molecule under coherent $J$-coupling evolution but coherence with the labile proton is constantly lost due to exchange, decreasing the overall signal lifetime. 
We assign the broad (relative to non-exchanging systems) $J$-spectra in \cref{fig:1DJspectra} to this effect, and the difference in linewidths between ethanol and acetic acid to different kinetics for the hydrogen-exchange reaction.

\begin{table}[h]
	\centering
\begin{tabular}{@{}lrrr@{}} \toprule
	$J$ (Hz) & 1-\carbon-EtOH & 2-\carbon-EtOH &\carbon\!$_2$-AcOH \\ \midrule
	\JoneCH & 140.852(1) & 125.2572(9) & 129.5041(3)\\
	\JoneCC &  ---& --- &56.7928(8) \\
	\JtwoCH & -4.700(3) & -2.317(3) &-6.7335(3) \\ 
	\JthreeHH & 7.049(3) &  7.032(2)   &--- \\  \bottomrule
\end{tabular}
\caption{Measured $J$-coupling constants for the two single labeled \carbon isotopomers of ethanol (EtOH) and the doubly \carbon labeled isotopomer of acetic acid (AcOH). The number in parenthesis denotes the 95\% confidence interval of the fit.}
\label{tab:exptJCouplingsEthanol}
\end{table}

\section*{Appendix: Multiple-Quantum Excitation and Filtering at Zero Field}

It would be desirable to selectively excite or at least detect only those coherences for which $\Delta F = n$, where $n$ is a chosen coherence order. This would assist with assigning quantum numbers, as the largest available $\Delta F$ value depends on spin-manifold, and it would also simplify the resulting spectra, since the number of transitions decreases rapidly with $n$. 
Existing high-field schemes for selective excitation \cite{Warren1979,Warren1980} and filtration \cite{Shaka1983} based on $\Delta m$ rely on the fact that coherence operators in high field have well defined symmetry with respect to rotations about the z-axis - a high-field $n$-th order coherence is invariant with respect to a rotation of $2\pi/n$ about the z-axis (we note that similar work has been done in an atomic-physics context \cite{Yashchuk2003}). 
Meanwhile, at zero-field, transitions in general involve changes in the total angular momentum of the state, not the projection on some axis. 
A zero-field $n$-order transition operator, $\ket{F,m_F}\bra{F-n,m_F}$ may be decomposed into spherical tensors \cite{VMK}. 
The lowest rank a given transition operator may contain is equal to $\Delta F$ and the highest is $2F-\Delta F$. 
In order to obtain simplified spectra one could imagine using a phase-cycling scheme similar to Spherical Tensor Analysis (STA)\cite{Beek24} to filter out components of rank lower than $n$ and higher than $2F-n$, which would increase the relative intensity of $n$-quantum transitions. 
However, dipole allowed transitions with $\Delta F = 0,\pm1$ may also contain components with ranks equal to or higher than $n$, so such a filtering process would therefore not suppress allowed transitions entirely. 

This reason for selective filtering of MQ-ZULF coherences being challenging also prevents selective excitation \cite{Warren1979,Warren1980}. 
Assuming that one can implement an ideal average Hamiltonian of rank $n$, the propagator to second order in time would still contain all ranks from $0$ to $2n$, so beyond the short time limit such a Hamiltonian would still excite the entire spectrum.

Indeed, from the point of view of spectral simplification it might be more fruitful to use an isotropic filtration phase-cycle as described by \cite{Pileio2008}, which retains only those signal components that are rank-0 during $t_1$ evolution. 
If applied to a zero-field experiment the resulting spectrum would consist only of those transitions for which $\Delta F = 0$, since only such coherence operators may contain rotationally invariant rank-0 components. 
There are only two such transitions in the case of \carbon\!$_2$-acetic acid (at $\sim$118\,Hz, and $\sim$264\,Hz), and consequently application of such a phase-cycle would result in a spectrum containing only two peaks instead of eight. 
In order to implement either isotropic filtering or STA one needs to be able to effect global rotations of the entire spin-system with arbitrary angles -- fortunately, this problem has recently been solved for zero-field systems \cite{Jiang2017}. 
However, the application of a multi-step phase cycle to an indirectly detected spectrum would be a practical challenge, particularly in terms of total experiment duration.

\begin{figure}[t!]
	\centering
	
	\includegraphics[]{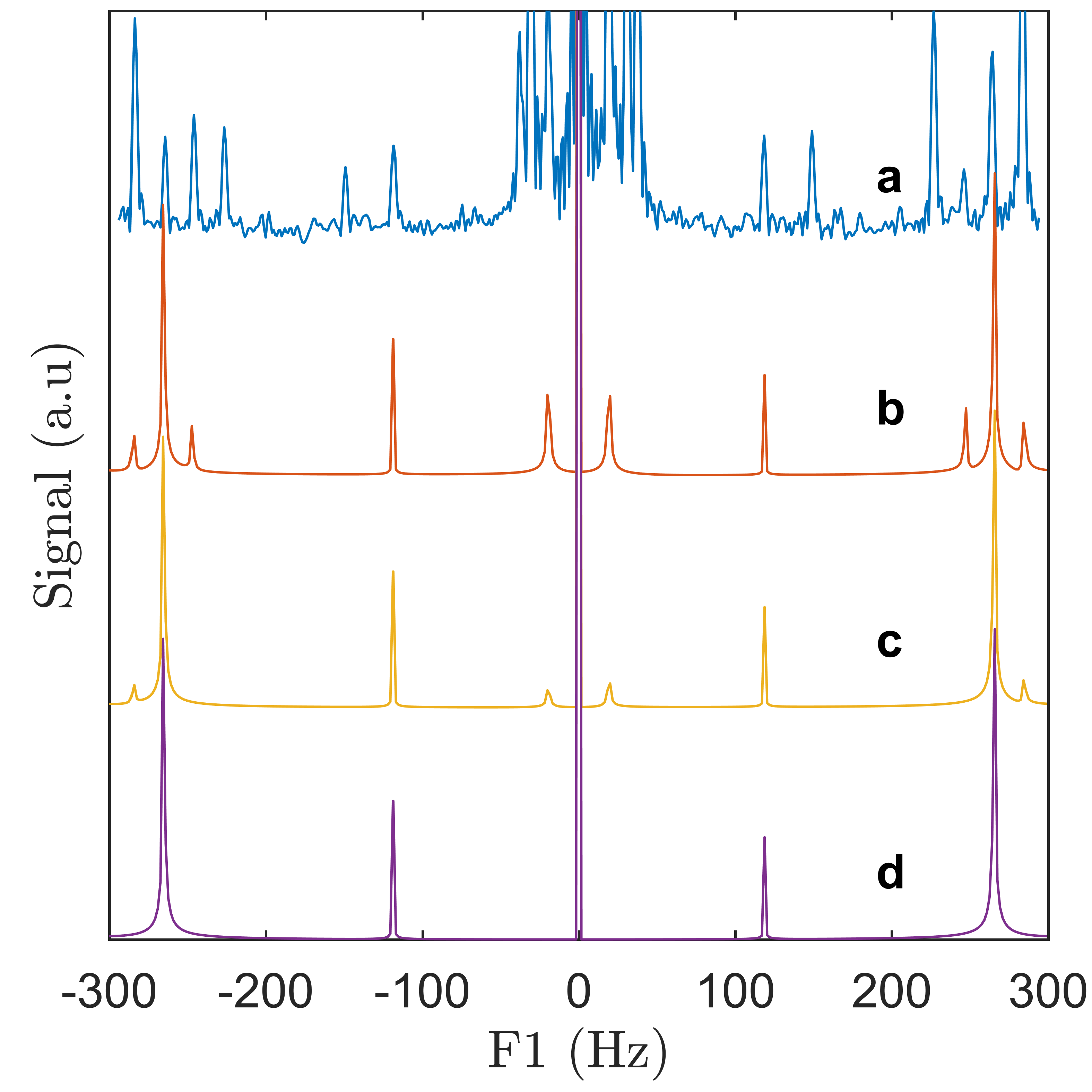}
	\caption{(a) Experimental MQ-ZF spectrum acquired using a rank-2 excitation Hamiltonian. (b) Simulated spectra showing the effect of a tetrahedral isotropic-filtering phase cycle. (c) Simulated octahedral phase cycle. (d) Simulated icosahedral phase cycle.}
	\label{fig:DQexcitationSpectra}
\end{figure}

The result of attempted rank-selective excitation is shown in \cref{fig:DQexcitationSpectra}a which is a MQ-ZULF spectrum of \carbon$_2$ acetic acid, but acquired using the effective excitation Hamiltonian
\begin{multline}
\bar{H}_{DQ}^1 =  2\pi(\ ^1J_{\mathrm{CH}}/2)(K_+S_{A,+}  K_-S_{A,-})\ + \\ 2\pi(\ ^3J_{\mathrm{CH}}/2)(K_+S_{B,+}  K_-S_{B,-}),
\end{multline}
which is implemented using a series of DC pulses $(xy\bar{y}\bar{x})_n$, where $x$ and $y$ denote $\pi$ pulses on carbon and the overbars denote reversals of the direction of the pulse. Such a `double-quantum' Hamiltonian would excite only even orders of coherence in a high-field system. The spectrum in \cref{fig:DQexcitationSpectra}a is the same kind of data as in \cref{fig:fullAcOH_MultiPolar}, but all the individual spectra have been added together. While the $n=2$ transition is successfully excited, it is clear that no selectivity has been gained over excitation with two-pulses and a delay as in \cref{fig:fullAcOH_MultiPolar2}, consistent with the above arguments about the difficulty of achieving selective excitation.

To demonstrate the potential benefits of isotropic coherence filtration for zero-field spectroscopy, Figs.\,\ref{fig:DQexcitationSpectra}b-d show simulated MQ-ZULF spectra resulting from the application of an isotropic-filter phase cycle, which averages to zero those components of the coherence operators that during $t_1$ have ranks up to 2 (b), 3 (c), and 4 (d) \cite{Pileio2008}. 
The largest rank supported by any coherence operator in this spin system is four, and consequently there is no reason to average higher-rank terms. 
The peak at $\sim$149\,Hz corresponds to an $F=1/2\rightarrow F=3/2$ transition and the corresponding coherence operator may only contain ranks 1 and 2. 
This peak disappears completely when operators up to rank-2 are filtered out using a 12-step cycle. 
The $n=2$ coherence may only contain ranks 2 and 3, and disappears when operators up to rank-3 are averaged out using a 24-step cycle, and when operators up to rank-4 are filtered out with a 60-step cycle only the two peaks remain which correspond to $\Delta F = 0$ transitions. 
While we are not yet able to selectively filter transitions based on $n$ for $n>0$, it seems isotropic filtering would allow spectral simplification by retaining only those transitions for which $n=0$. 
Note that the filtering has to be performed on the indirectly detected dimension, since only rank-1 components of the coherence operators correspond to observable magnetization.

We note in passing that, just as in high field, the rank of a coherence operator reports on the minimum number of correlated spins in the corresponding state. 
In high field this has been used to investigate spin cluster size and perform many-body physics experiments. 
The number of phase-cycle steps required to suppress a zero-field transition (plotted vs. preparation time) could similarly be used to monitor information transport in spin clusters connected with a Heisenberg-like isotropic Hamiltonian.

\end{document}